\documentclass[aps, twocolumn, reprint]{revtex4-2}

\usepackage{xcolor,color}
\usepackage{amsmath,amssymb}
\usepackage{mathtools}
\usepackage{upgreek}
\usepackage{graphicx}
\usepackage{tikz}

\usepackage[absolute]{textpos}

\usepackage[colorlinks, allcolors=blue]{hyperref}
\usepackage[figure, figure*]{hypcap}

\usetikzlibrary{calc}
\usetikzlibrary{decorations.pathmorphing}

\graphicspath{{fig}}

\newcommand{\Tr}[1]{\mathrm{Tr}#1}

\newcommand{\bra}[1]{\left\langle#1\right|}
\newcommand{\ket}[1]{\left|#1\right\rangle}
\newcommand{\braket}[2]{\left\langle#1\middle|#2\right\rangle}
\newcommand{\ketbra}[2]{\left|#1\right\rangle\left\langle#2\right|}

\newcommand{\Rmu}{\ensuremath{R_\mu}}
\newcommand{\Rnu}{\ensuremath{R_\nu}}
\newcommand{\Rnull}{\ensuremath{R^0}}

\newcommand{\Ntwo}{N\textsubscript2}

\newcommand{\Nitrogen}[6]{
\begin{tikzpicture}[baseline=(x.base)]

\coordinate (x) at (0,0);
\draw[thick] (0,-0.5,0) -- (0,0.5,0);

\draw[->,red,thick] (0,-0.5,0) -- ++(0.5*#1,0.5*#2,1*#3);
\draw[->,red,thick] (0,+0.5,0) -- ++(0.5*#4,0.5*#5,1*#6);

\shade[ball color = blue!40] (0,0.5) circle (0.2);
\shade[ball color = blue!40] (0,-0.5) circle (0.2);

\end{tikzpicture}
}

\def\address{%
    Institut f\"ur Theoretische Physik,
    Bremen Center for Computational Materials Science,
    and MAPEX Center for Materials and Processes,
    Otto-Hahn-Allee 1,
    Universit\"at Bremen,
    D-28359 Bremen,
    Germany}

\begin{document}

\title{Downfolding approaches to electron-ion coupling: Constrained density-functional perturbation theory for molecules}

\author{Erik G. C. P. van Loon}
\author{Jan Berges}
\author{Tim O. Wehling}
\affiliation\address

\begin{abstract}
Constrained electronic-structure theories enable the construction of effective low-energy models consisting of partially dressed particles. However, the interpretation and physical content of these theories is not straightforward. Here, we carefully explore the properties of downfolding theories for electron-ion problems, in particular constrained density-functional perturbation theory (cDFPT). We show that the dipole selection rules determine whether the partially dressed phonons satisfy Goldstone's theorem, and we prove that electronic screening always lowers the phonon frequencies. We illustrate the theory with cDFPT calculations for minimal example systems: the nitrogen and benzene molecule as well as graphene.
\end{abstract}

\maketitle

\begin{textblock*}{\paperwidth}(0mm, \paperheight-18mm)
    \centering
    \small
    Published in \href{https://doi.org/10.1103/PhysRevB.103.205103}{Phys.\@ Rev.\@ B \textbf{103}, 205103 (2021)} \copyright~American Physical Society
\end{textblock*}

\section{Introduction}

Electrons and nuclei together determine the microscopic properties of materials and molecules. The quantitative understanding of this interplay is a formidable task, since it is a many-body problem involving a large number of quantum particles.

\emph{Ab initio}--derived low-energy models are a way to address this problem~\cite{Imada10}. The electronic structure is divided into high- and low-energy states. The high-energy states are integrated out, leaving an effective low-energy model. The ``bare'' particles that enter the low-energy model are in fact the partially dressed particles of the full system. These low-energy degrees of freedom can subsequently be analyzed in more detail. For example, a detailed and computationally expensive treatment of electronic~\cite{Kotliar06} and electron-ion~\cite{Giovannetti14,Nomura15} correlations as well as complex dynamical and nonequilibrium phenomena is often only feasible for the low-energy model.

\emph{Ab initio}--based low-energy models also form the basis for including environmental effects like screening~\cite{Rosner16} and hybridization~\cite{Hall19}. This two-step approach has the big advantage that the changes in the environment only enter the second step of the evaluation, which can be substantially cheaper to evaluate than the full calculation.

One way for establishing low-energy models based on \emph{ab initio} calculations are the so-called ``constrained'' methods.
In these methods, the low-energy degrees of freedom are frozen, so that the effective interaction is screened only by processes involving high-energy electrons.
In this way, the constrained density-functional perturbation theory (cDFPT)~\cite{Nomura15} creates a low-energy model consisting of partially screened phonons, low-energy electrons, and an electron-phonon interaction. These three quantities can all be extracted from \emph{ab initio} calculations.
Similar constrained theories exist for the electron-electron interaction~\cite{Anisimov91}, in particular the constrained random-phase approximation~\cite{Aryasetiawan04}. Together, these approaches have allowed for the investigation of the combined effect of electron-electron and electron-phonon interactions, e.g., in the fullerides~\cite{Gunnarsson97,Capone02,Nomura15,Nomura15b,Nomura16,Nomura16b}. The cDFPT has been applied to several materials with electron-phonon coupling~\cite{Arita17,Hall19,Novko20,Berges20}. While these works provide answers to the physical problems they address, they naturally also raise questions regarding potential perks and pitfalls of the theory itself.

Here, our aim is to improve the understanding of the general structure and properties of downfolding theories for electron-ion problems in general and of cDFPT in particular.
We show that Goldstone's theorem does not generally apply to partially dressed phonons but that symmetry-based selection rules allow us to construct electronic target spaces that satisfy Goldstone's theorem. We also show that the electronic screening reduces the phonon frequencies and consider the basis transformation between bare and dressed phonons, i.e., harmonic mode-mode coupling.

To illustrate these findings, we have calculated the electron-ion coupling in small molecules. The equations of cDFPT are matrix relations in terms of both electronic and vibronic/phononic modes. In a crystalline solid, this means that all objects carry momentum labels in addition to their mode label and accurate calculations require a dense momentum grid. On the other hand, in molecules there is no momentum, the Hilbert space is finite, and all calculations are substantially easier. This allows us to elucidate important aspects of cDFPT in unprecedented detail.

The setup of this paper is as follows: First, we construct a general framework for the calculation of partially and fully dressed phonon properties within the Born-Oppenheimer approximation. We then show how cDFPT fits in this general framework and prove several properties of cDFPT\@. Subsequently, we illustrate the theory with numerical calculations for nitrogen, benzene, and, as a brief outlook towards periodic systems, graphene.

\section{Method}

Let us start with some remarks on terminology: We will generally use the term phonons for the ionic displacement eigenmodes, even for molecules, where one might also call them vibrons. We will be calculating the classical energy landscape corresponding to these displacements. We use the term ions to denote the nuclei and any core electrons that are fixed to the nuclei in the electronic-structure calculations. We set $\hbar=1$ and measure all energies and frequencies in eV.

We follow a variational approach to the electron-ion coupling~\cite{Gonze92,Putrino00,Refson06} and the screening of phonons in general, and we only specify density-functional theory (DFT) at the end. We show that the relations between full density-functional perturbation theory (DFPT) and cDFPT are particularly clear in this variational description.

\subsection{\emph{Ab initio} electronic structure}

The starting point for our analysis is the Born-Oppenheimer approximation: ionic and electronic degrees of freedom are formally separated. The electronic coordinates are supposed to be much faster, so that we can assume that they are always relaxed into an instantaneous ground state corresponding to a specific ionic configuration.

For a system consisting of $N$ ions, there are $3N$ ionic coordinates \Rmu. The electronic degrees of freedom are written as $\psi \in \Omega$ where $\Omega$ is a sufficiently smooth manifold. $\Omega$ could consist, e.g., of wave functions or density matrices, depending on the electronic-structure method that is used. We assume that there is an energy functional $\mathcal{E}(\psi;\Rmu)$ and that this functional is sufficiently smooth to calculate all necessary derivatives. The energy $E(\Rmu)$ corresponding to a fixed set of ionic coordinates \Rmu\ is found as the minimum of the functional $\mathcal{E}$ with respect to $\psi$: $E(\Rmu) = \min_{\psi \in \Omega} \mathcal{E}(\psi;\Rmu)$. The electronic configuration minimizing~\footnote{We assume that the minimum is unique.} the energy $\psi^\ast(\Rmu)$, implicitly defined by $E(\Rmu) = \mathcal{E}(\psi^\ast;\Rmu)$, depends on \Rmu. Therefore, a variation of \Rmu\ has two effects on the energy $E(\Rmu)$: explicitly, and implicitly via $\psi^\ast(\Rmu)$. The latter describes the \emph{electronic screening} of ionic displacements and is the main purpose of our investigations.

\subsection{Force}

The first derivative of the energy is the force  $F$. This is a $3N$-vector, containing the force on every ion as 3-vectors. The force corresponding to a specific ionic configuration $\Rnull$ is
\begin{align}
-F_\mu(\Rnull) &= \frac{d E}{d\Rmu} \Big|_{\Rnull} \\
&= \frac{\partial \mathcal{E}}{\partial \Rmu} + \frac{\partial \mathcal{E}}{\partial \psi} \frac{\partial \psi^\ast}{\partial \Rmu} \\
&=\frac{\partial \mathcal{E}}{\partial \Rmu} + 0 \cdot \frac{\partial \psi^\ast}{\partial \Rmu}. \label{eq:force}
\end{align}
The last line follows since $\psi^\ast(\Rnull)$ is the \emph{minimum} of $\mathcal{E}$.
This equation shows that the force can be obtained from $\psi^\ast(\Rnull)$ without having to take into account changes in the electronic configuration. This is a manifestation of the $2n+1$ theorem in electronic-structure theory~\cite{Gonze89,Gonze95,Gonze95b}.

For establishing low-energy models for a given ionic configuration \Rnull, we will use \emph{constrained} theories. They restrict the electronic configuration space to some subspace $\Omega'(\Rnull) \subset \Omega$, with $\psi^\ast(\Rnull)\in\Omega'$. This defines a new energy $E'(\Rmu)=\min_{\psi'\in\Omega'} \mathcal{E}(\psi';\Rmu)$ and a $\psi'^\ast(\Rmu)$ with $\mathcal{E}(\psi'^\ast(\Rmu);\Rmu)\equiv E'(\Rmu)$.
Since the constrained variational space is smaller, the following relations hold: \begin{align}
E(\Rmu) &\leq E'(\Rmu), \label{eq:energy:constrained}\\
E(\Rnull) &= E'(\Rnull), \label{eq:equalenergy} \\
-F'_\mu(\Rnull)&=\frac{dE'}{d\Rmu}\Big|_{\Rnull} \\
&= \frac{\partial \mathcal{E}}{\partial \Rmu} + 0 \cdot \frac{\partial \psi'^\ast}{\partial \Rmu}\\&=-F_\mu(\Rnull). \label{eq:equalforces}
\end{align}
The constrained theory gives the same forces at \Rnull, since the change in $\psi$ does not enter the equation for the force.

\subsection{Phonons}
\label{sec:phonons}

Important information about the ionic degrees of freedom is contained in the second derivative of the energy, a $3N\times 3N$ matrix. The eigenmodes of the dynamical matrix $\hat{\omega}^2$, with $\sqrt{m_\mu m_\nu} \hat{\omega}^2_{\mu \nu} = d^2 E/d\Rmu d\Rnu$, where $m_\mu$ are atomic masses, are called phonons and the associated eigenvalues are the phonon frequencies/energies.
Unlike the forces, the dynamical matrix is different in the constrained theory,
\begin{align}
\hat{\omega}^2_{\mu \nu} &= \frac{1}{\sqrt{m_\mu m_\nu}}\frac{d^2 E}{d\Rmu d\Rnu},  \\
\hat{\omega}'^2_{\mu \nu} &= \frac{1}{\sqrt{m_\mu m_\nu}}\frac{d^2 E'}{d\Rmu d\Rnu},  \\
\hat{\omega}^2 &\leq \hat{\omega}'^2. \label{eq:energydifference}
\end{align}
The last inequality follows from Eqs.~\eqref{eq:energy:constrained}, \eqref{eq:equalenergy}, and \eqref{eq:equalforces}. The inequality should be understood in the usual way for (symmetric) matrices: the difference $\Delta \hat{\omega}^2 = \hat{\omega}'^2 - \hat{\omega}^2$ is a positive-definite matrix.

Equation~\eqref{eq:energydifference} shows that the phonons in the constrained theory will always have a higher energy than in the full theory. The constraints prevent the electrons from completely moving along with the ions, thus increasing the energy cost of ionic displacements.

\subsection{Goldstone's theorem}
\label{sec:goldstone}

Goldstone's theorem states that every spontaneously broken continuous symmetry creates a massless ($\omega=0$) bosonic excitation.
In electronic-structure theory, the ions break the three continuous translation symmetries, creating three \emph{acoustic} phonon modes. In addition, molecules can have spontaneously broken rotation symmetries and corresponding massless modes.

Goldstone's theorem for phonons easily follows from our construction of the dynamical matrix. Assume that there is a continuous symmetry $T_\lambda$ parametrized by a real number $\lambda$. The total energy is invariant under this symmetry, $E(T_\lambda \Rmu) = E(\Rmu)$, and it is therefore possible to construct a representation of $T_\lambda$ acting on the $\psi\in\Omega$, with $\mathcal{E}(T_\lambda\psi;T_\lambda\Rmu)=\mathcal{E}(\psi;\Rmu)$. Concretely, a translation that acts on both the ionic coordinates \Rmu\ and the electronic configuration $\psi$ will leave the total energy unchanged.

Since $T_\lambda$ is a continuous symmetry, we can take $\lambda$ small and write $T_\lambda \Rmu \approx \Rmu+\lambda \, \Delta \Rmu$. The translation defines a direction in displacement space and the energy is constant in this direction, $E(\Rmu)=E(\Rmu+\lambda \, \Delta \Rmu)$. This displacement is thus an eigenmode of the dynamical matrix, with eigenvalue (energy) zero.

This argument does not transfer to the constrained theory~\cite{Nomura15}. The constraints can break the continuous symmetry explicitly: $T_\lambda \psi \notin \Omega'(\Rnull)$. In that case, the electronic configuration cannot completely move along with the translation symmetry due to the constraints and $E'(T_\lambda \Rnull) \neq E'(\Rnull)=E(\Rnull)=E(T_\lambda \Rnull)$. Due to this last equality, we conclude $E'(T_\lambda \Rnull) > E'(\Rnull)$. However, this inequality is not yet sufficient to conclude that the Goldstone modes will always acquire a finite energy in the constrained theory: It is still possible that constrained and full theory agree to order $\delta R^2$ and only differ at order $\delta R^4$ (or higher). In that case, the partially dressed phonons will still satisfy Goldstone's theorem. For cDFPT, we show in Sec.~\ref{eq:savinggoldstone} that dipole selection rules can be used to construct a $\Omega'$ that guarantees that the uniform translation modes stay massless.

\subsection{Bare and dressed frequencies}

This brings us to the relation between the bare and dressed dynamical matrices. Both are second derivatives of the energy with respect to the displacement, with the subtlety that the electronic configuration adjusts to the displacement. This electronic response is where the difference between the full and constrained theory (i.e., DFPT and cDFPT) originates from. It is useful to consider this point in detail. We measure the atomic displacements $\delta R$ with respect to some initial ionic positions $\Rnull$ with electronic configuration $\psi^0=\psi^\ast(\Rnull)$. To determine the second derivative, it is sufficient to Taylor expand the energy functional to second order in $\delta R$ and $\delta\psi$. Summation over repeated indices is implied and we use Latin indices $a,b$ for the electronic degrees of freedom~\footnote{$\Omega$ is a manifold and thus locally equivalent to a vector space, so the derivative $\delta\psi$ can be decomposed into components.}.
\begin{widetext}
\begin{align}
\mathcal{E}(\psi;\Rnull+\delta R) - \mathcal{E}(\psi^0;\Rnull) &=
\frac{\partial \mathcal{E}}{\partial \Rmu} \delta\Rmu
+
\frac{1}{2} \delta \psi_a \frac{\partial^2 \mathcal{E}}{\partial \psi_a \partial \psi_b}  \delta\psi_b
+
\frac{1}{2} \delta \Rmu \frac{\partial^2 \mathcal{E}}{\partial \Rmu \partial \Rnu} \delta\Rnu
+
\delta \psi_a \frac{\partial^2 \mathcal{E}}{\partial \psi_a \partial \Rmu} \delta \Rmu.
\label{eq:energy:taylor}
\end{align}
\end{widetext}
There is no first-order contribution in $\delta \psi$, as we saw when calculating the force.
For a given displacement $\delta R$, there is a physical (minimal energy) solution $\psi^\ast(\delta R)$ defined by $\partial \mathcal{E}(\psi^\ast(\delta R);\Rmu)/\partial \psi_a=0$. At small $\delta \Rmu$, $\delta \psi_b^\ast(\delta R)$ is linear, with
\begin{align}
0&= \frac{\partial^2 \mathcal{E}}{\partial \psi_a \partial \psi_b} \delta \psi_b + \frac{\partial^2 \mathcal{E}}{\partial \psi_a \partial \Rmu} \delta \Rmu,
\\
\delta \psi_b^\ast(\delta R) &= -
\left(
\frac{\partial^2 \mathcal{E}}{\partial \psi_a \partial \psi_b}
\right)^{-1}
\left(
\frac{\partial^2 \mathcal{E}}{\partial \psi_a \partial \Rmu}
\right) \delta\Rmu \\
&\equiv
-\mathcal{E}^{-1}_{\psi_a\psi_b} \mathcal{E}_{\psi_a \Rmu} \delta\Rmu.
\end{align}
Here, we have introduced the subscript partial-derivative notation $\mathcal{E}_{x}=\partial \mathcal{E}/\partial x$, and the inverse is a matrix inversion.
We reinsert this result into the energy functional, Eq.~\eqref{eq:energy:taylor},
\begin{multline}
\mathcal{E}(\psi^\ast(\delta R);\Rnull+\delta R) - \mathcal{E}(\psi^0;\Rnull)
=
\mathcal{E}_{\Rmu} \delta\Rmu
\\
+
\frac{1}{2}\delta \Rmu\left[
\mathcal{E}_{\Rmu\Rnu}
-
\mathcal{E}_{\Rmu\psi_a} \mathcal{E}^{-1}_{\psi_a\psi_b} \mathcal{E}_{\psi_b \Rnu}
\right] \delta \Rnu.
\end{multline}
The term in brackets determines the interatomic force constants,
\begin{align}
\sqrt{m_\mu m_\nu}\hat{\omega}_{\mu\nu}^2 &= \frac{d^2 E}{d\Rmu d\Rnu}  \\
&= \frac{d^2 \mathcal{E}(\psi^\ast(R);R)}{d\Rmu d\Rnu} \\
&= \mathcal{E}_{\Rmu\Rnu}
-\mathcal{E}_{\Rmu\psi_a} \mathcal{E}^{-1}_{\psi_a\psi_b} \mathcal{E}_{ \Rnu\psi_b}. \label{eq:dynamicalmatrix}
\end{align}
The first term here is the bare energy cost of displacing ions with fixed electronic configuration, the second term represents the screening by the electrons. Note that the second derivative $\mathcal{E}_{\psi_a\psi_b}$ is the Hessian on the electronic manifold and summation over the electronic degrees of freedom is implied.
Since $\psi^\ast$ is a \emph{local minimum} of the energy functional, the electronic Hessian $\mathcal{E}_{\psi\psi}$ is positive definite and the second term in Eq.~\eqref{eq:dynamicalmatrix} can be interpreted as an inner product $\langle \mathcal{E}_{\Rmu\psi} , \mathcal{E}_{\Rnu\psi} \rangle_{\mathcal{E}^{-1}_{\psi\psi}}$. This shows that the $\mu=\nu$ part of this term is always negative (due to the $-1$ in front of the inner product) and therefore \emph{reduces} the eigenvalues of $\hat{\omega}^2$. Screening reduces the phonon energies.

The relation between dynamical matrices of the full and the constrained theory is particularly clear in Eq.~\eqref{eq:dynamicalmatrix}. The second term contains an implicit internal sum over the dimensions of the electronic manifold, and in the constrained theory this sum is restricted to the submanifold.
It can be useful to analyze this sum term by term, i.e., to perform so-called fluctuation diagnostics~\cite{Gunnarsson15,Berges20}.

\subsection{Phonons and density-functional theory}

Many flavors of \emph{ab initio} calculations exist, specified by their electronic coordinate space $\Omega$ and energy functional $\mathcal{E}$.
The relations and proofs given so far do not depend on the precise choice of $\mathcal{E}$---as long as a single variational functional is used consistently---although there will of course be quantitative differences.
In the remainder of this paper, we restrict ourselves to DFT, but one could also apply these considerations to, e.g., Hartree-Fock or variational Monte Carlo.

The determination of phonons and the electron-phonon coupling from DFT is done via DFPT\@. Here, we will only state the most relevant formulas; more detailed derivations are found elsewhere~\cite{Baroni01,Nomura15}.

In the framework of DFT, the energy is a functional $\mathcal{E}(\rho; R)$ of the electronic density $\rho(r)$ and ionic coordinates $R$.
The electronic properties are most easily expressed using the Kohn-Sham basis $\{\ket{m}\}$. In this basis, the electronic density is represented by a density matrix $\rho$.
In other words, using $m,n$ to denote the electronic levels, the connection to the previous sections is $\psi_a \equiv \rho_{mn}$. Every degree of freedom $a$ in the electronic manifold corresponds to an electronic transition $(mn)$, and when there are $N_\text{KS}$ Kohn-Sham states, the electronic manifold $\Omega$ consists of $N_\text{KS} \times N_\text{KS}$ matrices and the number of electronic degrees of freedom $N_\text{el}$ satisfies $N_\text{el} \leq N_\text{KS}\times N_\text{KS}$, where the inequality follows since not every matrix is also a valid density matrix. See Appendix~\ref{app:kohnsham} for a minimal example.
The matrix structure is necessary since ionic displacements generally lead not just to changes in Kohn-Sham energies but also to the hybridization of Kohn-Sham orbitals.

The interatomic force constants can be expressed in terms of the electronic \emph{ground-state} density $\rho(r; R)$ for given $R$ as
\begin{multline}
    \label{eq:IFC_DFT}
    \frac{d^2 \mathcal E(\rho; R)}{d \Rmu d \Rnu }
    = \int d^3 r \,
    \frac{\partial V_\text{ext}(r; R)}{\partial \Rmu}
    \frac{\partial \rho(r; R)}{\partial \Rnu}
    \\
    + \int d^3 r \,
    \frac{\partial^2 V_\text{ext}(r; R)}{\partial \Rmu \partial \Rnu}
    \rho(r; R)
    + \mathcal E_{\Rmu \Rnu},
\end{multline}
where $V_\text{ext}$ is the bare external potential of an individual electron amid the ensemble of ions and $\mathcal E_{\Rmu \Rnu}$ is the second derivative of the ionic repulsion.

The change of the electronic density follows from the Kohn-Sham equations and reads
\begin{multline}
    \label{eq:drho}
    \partial \rho(r; R)
    = 2 \sum_{\mathclap{m, n \leq N_\text{KS}}}
    \frac
        {f(\epsilon_m) - f(\epsilon_n)}
        {\epsilon_m - \epsilon_n}
    \\
    \times
    \braket n r
    \braket r m
    \bra m
        \partial \hat V_\text{eff}(R)
    \ket n,
\end{multline}
where $V_\text{eff}$ is the dressed self-consistent potential and $\ket m$, $\ket n$ are Kohn-Sham eigenstates. Since $\partial V_\text{eff}$ depends on $\partial \rho$, this equation must be solved self-consistently. With that, we find the electronic contribution to the interatomic force constants,
\begin{multline}
    \label{eq:IFC_el}
    \int d^3 r \,
    \frac{\partial V_\text{ext}(r; R)}{\partial \Rmu}
    \frac{\partial \rho(r; R)}{\partial \Rnu}
    = 2 \sum_{\mathclap{m, n \leq N_\text{KS}}}
    \frac
        {f(\epsilon_m) - f(\epsilon_n)}
        {\epsilon_m - \epsilon_n}
    \\
    \times
    \bra n
        \frac{\partial \hat V_\text{ext}(R)}{\partial \Rmu}
    \ket m
    \bra m
        \frac{\partial \hat V_\text{eff}(R)}{\partial \Rnu}
    \ket n.
\end{multline}
This is nothing but the bare electronic susceptibility $\chi^0$ weighted with the bare and dressed (DFPT) deformation-potential matrix elements $d_{\mu n m} = \bra n \partial \hat V_\text{ext} / \partial \Rmu \ket m$ and $\tilde d_{\nu m n} = \bra m \partial \hat V_\text{eff} / \partial \Rnu \ket n$, respectively. The deformation potential is related to the electron-phonon coupling as appears in a Hamiltonian via $g = d / \sqrt{2 \omega' m}$.

The cDFPT is obtained by picking a target subspace $N'_\text{KS}$ and restricting the summations in Eqs.~\eqref{eq:drho} and \eqref{eq:IFC_el} by excluding summands where \emph{both} $m$ and $n$ lie in the target space. This yields the partially screened phonon frequencies.
The relation between the dynamical matrices in DFPT ($\hat{\omega}^2$) and cDFPT ($\hat{\omega}'^2$) is~\cite{Berges20} given by
\begin{multline}
    \sqrt{m_\mu m_\nu} \hat \omega^2_{\mu \nu}
    = \sqrt{m_\mu m_\nu} \hat \omega'^2_{\mu \nu}
    \\
    + 2 \sum_{\mathclap{m, n \in \text{target}}}
    d^\ast_{\mu m n}
    \frac
        {f(\epsilon_m) - f(\epsilon_n)}
        {\epsilon_m - \epsilon_n}
    \tilde d_{\nu m n}.
\end{multline}

The cDFPT formulation presented here seems to deviate from the general result Eq.~\eqref{eq:dynamicalmatrix}. In the latter, both ends of the self-energy diagram have the same electron-phonon vertex, which presently would read $\mathcal{E}_{R_\mu \rho_{mn}}$. These are connected by a susceptibility $\epsilon^{-1}_{\rho_{ab}\rho_{cd}}$, which is a $(N_\text{el}\times N_\text{el})\times (N_\text{el}\times N_\text{el})$ matrix. Here, the full susceptibility is replaced by the bare susceptibility and the electronic interactions are absorbed into one of the vertices, as in Fig.~\ref{fig:vertex}. The advantage of this approach is that the bare susceptibility is a diagonal $(N_\text{el}\times N_\text{el})\times (N_\text{el}\times N_\text{el})$ matrix and it is sufficient to only consider the $N_\text{el}\times N_\text{el}$ diagonal elements. The figures in this manuscript only show these $N_\text{el}\times N_\text{el}$ matrix elements.

\subsection{Saving Goldstone's theorem via selection rules}
\label{eq:savinggoldstone}

The partially dressed cDFPT phonons are not guaranteed to satisfy Goldstone's theorem. However, here we show that suitably chosen target spaces will guarantee massless phonons corresponding to uniform translations.

First, let us consider systems with an inversion symmetry, i.e., invariance under $x\mapsto -x$. For uniform translations, $\partial V/\partial R$ is odd under inversion. Thus, if all target-space orbitals are either even (\emph{gerade}) or odd (\emph{ungerade}) under inversion, then the displacement-potential matrix element $d=\bra{m}\partial V/\partial R \ket{n}$ is antisymmetric in total and therefore zero. In that case, there is no contribution to the phonon frequency from target-space electronic transitions and the partially dressed phonon has the same frequency as the fully screened mode, which is massless.

In fact, an even stronger dipole selection rule applies to uniform-translation phonons.
If a uniform translation $\lambda$ is performed on the nuclear coordinates, then the Kohn-Sham eigenfunctions are transformed by applying $\exp(i P\cdot \lambda)$, since the total momentum operator $P$ is the generator of uniform translations,
\begin{align}
 \ket{m_\lambda} = e^{iP\lambda} \ket{m} ,
\end{align}
and the Kohn-Sham eigenvalues are invariant under this transformation by symmetry.
The initial density matrix was
\begin{align}
 \rho = \sum_m f(\epsilon_m) \ketbra{m}{m},
\end{align}
and the density matrix after translation is
\begin{align}
 \rho_\lambda = \sum_m f(\epsilon_m)\ketbra{m_\lambda}{m_\lambda}.
\end{align}
We expand up to linear order in $\lambda$,
\begin{align}
 \rho_\lambda &= \sum_m f(\epsilon_m) (1+iP\lambda)\ketbra{m}{m} (1-iP\lambda), \notag \\
 \rho_\lambda-\rho &= \sum_m f(\epsilon_m) \big(iP\lambda\ketbra{m}{m}-\ketbra{m}{m}iP\lambda \big), \notag \\
 (\rho_\lambda-\rho)_{ab}
 &= i\lambda \big(f(\epsilon_b) -f(\epsilon_a) \big) \bra{a}P\ket{b}.
\end{align}

In the general analysis of Goldstone's theorem, we stated that the partially dressed phonon satisfies Goldstone's theorem if $\rho_\lambda=T_\lambda \rho \in \Omega'$ for small $\lambda$, where $\Omega'$ is the cDFPT submanifold consisting of density matrices with the target space subblock frozen to the initial value. In other words, $\rho_\lambda$ is part of the cDFPT submanifold as long as $(\rho_\lambda-\rho)_{ab}=0$ for all target states $\ket{a},\ket{b}$. The previous derivation shows that this is the case for uniform translation modes if $\bra{a}P\ket{b}$ is zero for any pair of target states $\ket{a},\ket{b}$, i.e., the Goldstone modes are preserved if there are no long-wavelength dipole-allowed~\footnote{The name dipole selection rule originates in the relation between the matrix elements of the momentum operator and the position operator. If $\hat{H}=\hat{p}^2/2m + V(\hat{r})$, then $[\hat{H},\hat{r}]=-i\hat{p}/m$ and $(E_a-E_b) \bra{a}\hat{r}\ket{b}=\bra{a} [\hat{H},\hat{r}]\ket{b}=-i/m \bra{a}\hat{p}\ket{b}$.
} transitions possible within the target space.

\section{Positivity of fluctuation diagnostics}
\label{sec:lowerenergy}

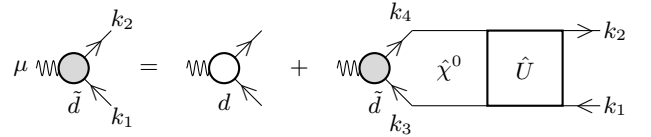
\begin{figure}
\begin{tikzpicture}

\draw (0.5,0.5) -- node[sloped,pos=0.4] {$>$} (0,0) ;
\draw (0.5,-0.5) -- node[sloped,pos=0.4] {$<$} (0,0) ;
\draw[decorate,decoration={snake,amplitude=1mm,segment length=1mm,post length=1mm}] (-0.5,0) -- (0,0) ;
\draw[thick,fill=gray!30] (0,0) circle (0.2);

\node at (0,-0.45) {$\tilde{d}$} ;
\node at (-0.7,0) {$\mu$} ;
\node at (+0.65,0.65) {$k_2$} ;
\node at (+0.65,-0.65) {$k_1$} ;
\begin{scope}[shift={(2,0)}]

\draw (0.5,0.5) -- node[sloped,pos=0.4] {$>$} (0,0) ;
\draw (0.5,-0.5) -- node[sloped,pos=0.4] {$<$} (0,0) ;
\draw[decorate,decoration={snake,amplitude=1mm,segment length=1mm,post length=1mm}] (-0.5,0) -- (0,0) ;
\draw[thick,fill=white] (0,0) circle (0.2);

\node at (0,-0.45) {$d$} ;
\end{scope}

\begin{scope}[shift={(4,0)}]

\draw (0.5,0.5) -- node[sloped,pos=0.4] {$>$} (0,0) ;
\draw (0.5,-0.5) -- node[sloped,pos=0.4] {$<$} (0,0) ;
\draw (0.5,0.5) -- (1.5,0.5) ;
\draw (0.5,-0.5) -- (1.5,-0.5) ;

\draw[decorate,decoration={snake,amplitude=1mm,segment length=1mm,post length=1mm}] (-0.5,0) -- (0,0) ;
\draw[thick,fill=gray!30] (0,0) circle (0.2);

\node at (1,0) {$\hat{\chi}^0$} ;
\node at (2,0) {$\hat{U}$} ;
\draw[thick] (1.5,0.5) -- ++(0,-1) -- ++(1,0) -- ++ (0,1) -- cycle;

\draw (3,0.5) -- node[sloped,pos=0.4] {$>$} (2.5,0.5) ;
\draw (3,-0.5) -- node[sloped,pos=0.4] {$<$} (2.5,-0.5) ;

\node at (0,-0.45) {$\tilde{d}$} ;
\node at (+0.35,0.75) {$k_4$} ;
\node at (+0.35,-0.75) {$k_3$} ;

\node at (+3.2,0.5) {$k_2$} ;
\node at (+3.2,-0.5) {$k_1$} ;

\end{scope}

\node at (1,0)  {=} ;
\node at (3,0)  {+} ;

\end{tikzpicture}
\caption{Relation between dressed and bare electron-phonon coupling.}
\label{fig:vertex}
\end{figure}

As we have seen in Eq.~\eqref{eq:energydifference}, dressed phonons have a lower energy than bare phonons, since electrons can move along to screen the ionic charges. We have also stated that $\Delta \hat{\omega}^2$ can be interpreted mathematically as an inner product of electron-phonon vertices, with the metric given by the electronic susceptibility.
Here, we will make this proof more explicit for cDFPT\@. This analysis shows that every contribution $\Delta \hat{\omega}^2_{\mu\mu,mn}$ in the fluctuation diagnostics is positive. This is helpful, since it means that no cancellations occur and the relative contribution of specific fluctuations can be quantified easily.

The proof is based on the relation~\cite{Nomura15,Giustino17,Hedin69} between $\tilde{d}$ and $d$, also illustrated in Fig.~\ref{fig:vertex},
 \begin{align}
  \tilde{d}_{\mu k_1k_2} = d_{\mu k_3k_4} \left[\frac{\mathbb{I}}{\mathbb{I}-\hat{\chi}^0 \hat{U} }\right]_{k_3k_4,k_1k_2\displaystyle,} \label{eq:proof:1}
 \end{align}
 where $k_i$ labels the electronic states,
\begin{align}
    \hat{\chi}^0_{k_1k_2,k_3k_4} = \delta_{k_1 k_3} \delta_{k_2 k_4} \frac{ f(\epsilon_{k_2})-f(\epsilon_{k_1}) }{\epsilon_{k_2}-\epsilon_{k_1}}
\end{align}
is the Lindhard bubble in the electronic eigenbasis, and $\hat{U}$ is the electronic interaction kernel~\footnote{This electronic interaction is present since we need to consider the \emph{self-consistent} response of the electronic system, see Ref.~\citenum{Nomura15} for more details. In the current proof, we only require that this denominator does not change the sign of the eigenvalues. In other words, the electronic system should be thermodynamically stable~\cite{Stoner39}.}.
We once again see that all elements have either two or four electronic state labels, which motivates us to consider these labels pairwise, as transitions.
If there are $N_\text{el}$ electronic states, then $d_\mu$ is a vector in $\mathbb{C}^{N_\text{el}^2}$ and both $\hat{\chi}$ and $\hat{U}$ are $N_\text{el}^2 \times N_\text{el}^2$ matrices (or rank-4 tensors)~\cite{Kaltak15}. The inverse in Eq.~\eqref{eq:proof:1} should be understood as a matrix inversion in this space and $\mathbb{I}$ is the identity matrix.

The phonon self-energy $\Delta \hat{\omega}^2_{\mu\nu}= (\hat{\omega}^2_\text{cDFPT}-\hat{\omega}^2_\text{DFPT})_{\mu\nu}$ is a matrix in phonon-branch space. The fluctuation diagnostics looks at the contribution coming from a pair of electronic states $k_1, k_2$, which we denote by $\Delta \hat{\omega}^2_{\mu\nu,k_1k_2}$.

Looking at a symmetrized combination of the contributions from $k_1$ and $k_2$, we find  \begin{align}
  & \sqrt{m_\mu m_\nu} \Delta \hat{\omega}^2_{\mu\nu,k_1k_2} \\
  &= -\tilde{d}_{\mu k_1k_2} \hat{\chi}^0_{k_1 k_2, k_1 k_2} d^*_{\nu k_1k_2} - d_{\mu k_1k_2} \hat{\chi}^0_{k_1 k_2, k_1 k_2} \tilde{d}^*_{\nu k_1k_2}  \\
  &= -\sum_{k_3k_4}  d_{\mu k_3 k_4} \, \left[\frac{\hat{\chi}^0}{\mathbb{I} -\hat{U}\hat{\chi}^0 }\right]_{\mathrlap{k_3k_4,k_1k_2}} \, d^*_{\nu k_1k_2} -\ldots\\
  &= -\sum_{\mathclap{k_3 k_4, k_5 k_6}} d_{\mu k_3k_4} \left[\frac{\hat{\chi}^0}{\mathbb{I} -\hat{U}\hat{\chi}^0 } \hat{P}^{k_1k_2} \right]_{\mathrlap{k_3k_4,k_5k_6}}  d^*_{\nu k_5k_6} \label{eq:proof:2}-\ldots \\
  &\equiv \langle d_{\mu}, \hat{P}^{k_1 k_2} d_{\nu} \rangle^{\chi} + \langle \hat{P}^{k_1 k_2} d_{\mu},  d_{\nu} \rangle^{\chi},
 \end{align}
 where we have introduced the projection onto $k_1 k_2$: $[\hat{P}^{k_1k_2}]_{k_3k_4,k_5k_6} = \delta_{k_3k_4,k_5k_6}\delta_{k_1k_2,k_3k_4}$. In the last line, we have introduced the bilinear form corresponding to the operator $\hat{\chi}$, with
 \begin{align}
  \hat{\chi} = -\frac{\hat{\chi}^0}{\mathbb{I} -\hat{U}\hat{\chi}^0 } .
 \end{align}
In Appendix~\ref{app:lemma}, we prove that $\hat{\chi}$ is a positive-definite symmetric real matrix, so that the corresponding bilinear form is an inner product. $\hat{\chi}$ can be understood as the susceptibility, and the fixed sign of $\hat{\chi}$ is then related to thermodynamic stability, as discussed below Eq.~\eqref{eq:dynamicalmatrix}. From this, our desired results follow directly, namely that $\Delta \hat{\omega}^2_{\mu\mu,k_1k_2} \geq 0$ and that $\Delta \hat{\omega}^2_{\mu\mu}= \sum_{k_1k_2} \Delta \hat{\omega}^2_{\mu\mu,k_1k_2} = \langle d_{\mu}, d_{\nu} \rangle^{\chi} / \sqrt{m_\mu m_\nu} \geq 0$.

In this proof, $k_1$ and $k_2$ label the electronic eigenstates. Here, these are the molecular orbitals. In lattice systems, momentum is a good quantum number and the electronic states are labeled by momentum $\mathbf{k}$ and a band index. The momentum analysis is simplified by setting $\mathbf{k}_2=\mathbf{k}_1+\mathbf{q}$ and by then observing that the entire equation is diagonal in $\mathbf{q}$.

\section{Computational details}
\label{sec:computation}

The \emph{ab initio} calculations in this paper are realized using \textsc{Quantum ESPRESSO}~\cite{Giannozzi09,Giannozzi17}. We apply the generalized gradient approximation (GGA) by Perdew, Burke, and Ernzerhof (PBE)~\cite{Perdew96,Perdew97} and optimized norm-conserving Vanderbilt pseudopotentials~\cite{Hamann13} from the \textsc{PseudoDojo} pseudopotential table~\cite{vanSetten18} at a kinetic-energy cutoff of 150\,Ry. Core electrons are incorporated into the pseudopotential. The electronic temperature is set to zero (no occupation smearing). Forces are minimized to below 1\,$\upmu$Ry/Bohr. Unwanted interactions between periodic images of the system are kept small using a unit-cell size of 15\,\AA{} in the respective directions. For the calculations of graphene in Appendix~\ref{app:graphene}, we employ the lattice constant $a = 2.46$\,\AA{} and sample the Brillouin zone with $32 \times 32$ $\mathbf{k}$ and $8 \times 8$ $\mathbf{q}$ points including $\Gamma$.

\section{Nitrogen molecule}

\begin{figure}
\includegraphics[scale=0.7]{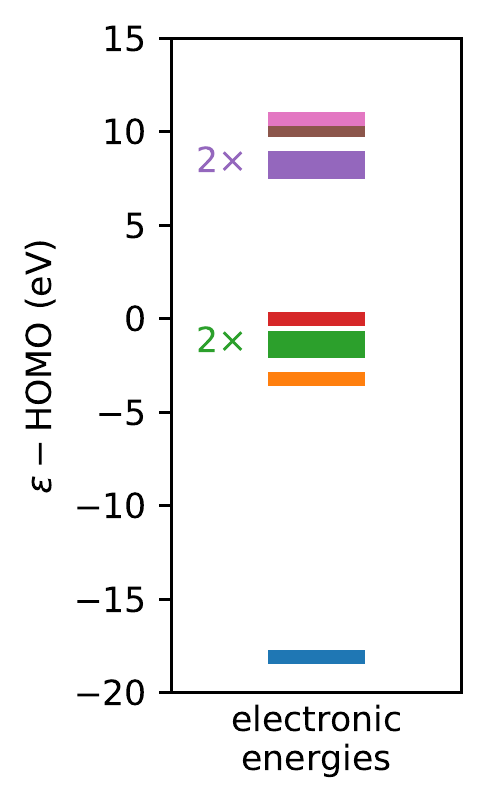}%
\includegraphics[scale=0.7]{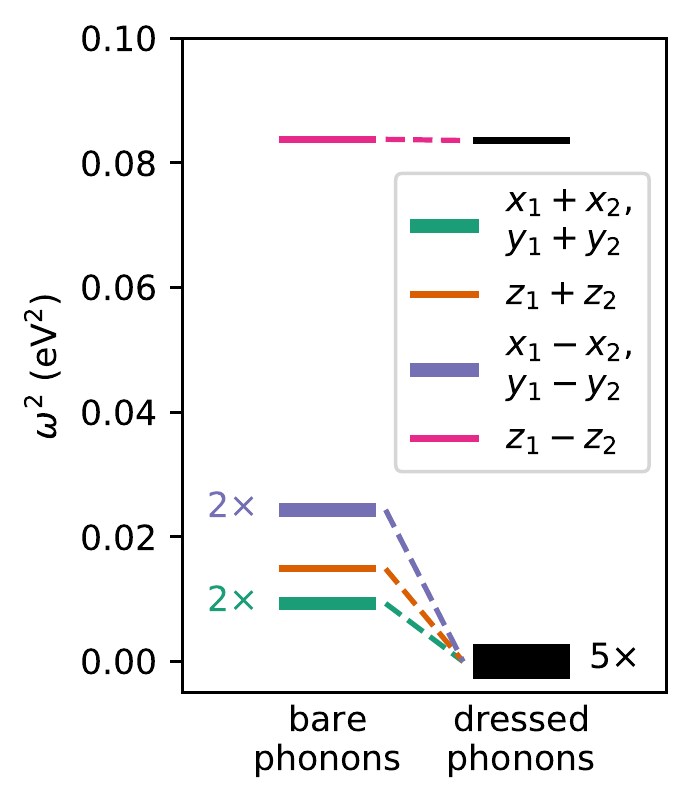}
\caption{Electronic and phononic spectra of an \Ntwo\ molecule. Note that the 1s electrons are incorporated into the pseudopotential in our calculations, so they are not shown here.}
\label{fig:N2:phonons}
\end{figure}

We consider an \Ntwo\ molecule, consisting of two identical N ions aligned along the $z$~axis. Explicitly, the atomic coordinates are $\Rmu = (0,0,-a/2,0,0,a/2)$, where $a = 1.1$\,\AA{} is the interionic distance. In our calculation, only the 1s electronic state is incorporated into the pseudopotential of the ion. The electronic energy levels are shown in Fig.~\ref{fig:N2:phonons}.

There are six atomic coordinates, so the dynamical matrix is a $6 \times 6$ matrix and there are six phonons.
Table~\ref{tab:N2} shows the displacements of the phonon eigenmodes~\footnote{Here, the phonon energies stand only for the second derivative of the classical potential-energy surface of the molecule in the Born-Oppenheimer approximation.}. These are the eigenmodes of both the cDFPT and the DFPT spectrum, so $\Delta \hat{\omega}^2_{\mu\nu}$ is a diagonal matrix.
For the calculation of the bare phonons, we have fixed all electronic energy levels of Fig.~\ref{fig:N2:phonons}. The resulting cDFPT and DFPT spectra are shown in Fig.~\ref{fig:N2:phonons}. Due to rotation symmetry in the $x$-$y$~plane, the eigenmodes with $x$ and $y$ displacements come in degenerate pairs.

\begin{table*}
\begin{tabular}{c c c c c c}
\Nitrogen{1}{0}{0}{1}{0}{0} &
\Nitrogen{0}{0}{1}{0}{0}{1} &
\Nitrogen{0}{1}{0}{0}{1}{0} &
\Nitrogen{1}{0}{0}{-1}{0}{0} &
\Nitrogen{0}{0}{1}{0}{0}{-1} &
\Nitrogen{0}{1}{0}{0}{-1}{0}
\\
$x$ translation
&
$y$ translation
&
$z$ translation
&
rotation around $y$
&
rotation around $x$
&
bond stretching
\\
$x_1 + x_2 $
&
$y_1 + y_2$
&
$z_1 + z_2 $
&
$x_1 - x_2 $
&
$y_1 - y_2$
&
$z_1 - z_2$
\\
(+1,0,0,+1,0,0)
&
(0,+1,0,0,+1,0)
&
(0,0,+1,0,0,+1)
&
(+1,0,0,$-1$,0,0)
&
(0,+1,0,0,$-1$,0)
&
(0,0,+1,0,0,$-1$)
\end{tabular}
\caption{The phonon eigenmodes of \Ntwo.}
\label{tab:N2}
\end{table*}

The five DFPT modes with vanishing energy are Goldstone modes: two spontaneously broken rotation symmetries (around $x$ and $y$) and three spontaneously broken translation symmetries ($x$, $y$, $z$).
In contrast, cDFPT  considers a system where the electronic density is fixed. These constraints have already explicitly broken the symmetries, see Sec.~\ref{sec:goldstone}, so there is no spontaneous symmetry breaking and there are no bare Goldstone modes. This is visible in the cDFPT spectrum; all five bare phonon modes have a finite energy.

As predicted in Sec.~\ref{sec:phonons}, the frequency of the dressed phonons in Fig.~\ref{fig:N2:phonons} is reduced compared to the bare modes, although this effect is small and hardly visible for the $z_1-z_2$ mode.

\subsection{Fluctuation diagnostics}
\label{sec:fluctuations}

\begin{figure}
\includegraphics{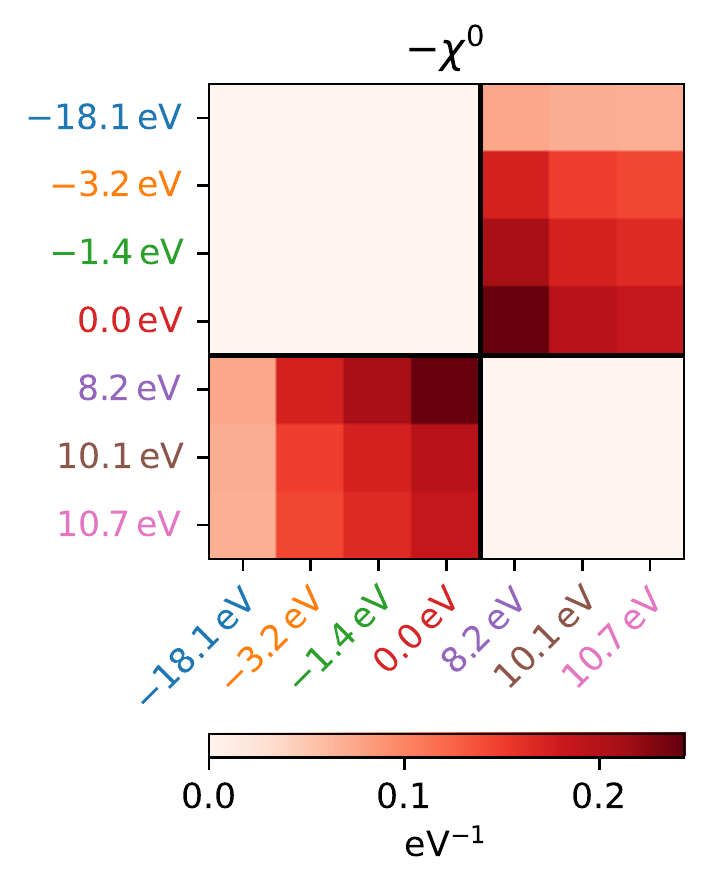}
\caption{Bare susceptibility $\chi^0$ of \Ntwo. The susceptibility is nonzero only for transitions across the Fermi level.}
\label{fig:N2:chi0}
\end{figure}

\begin{figure*}
\includegraphics[width=\linewidth]{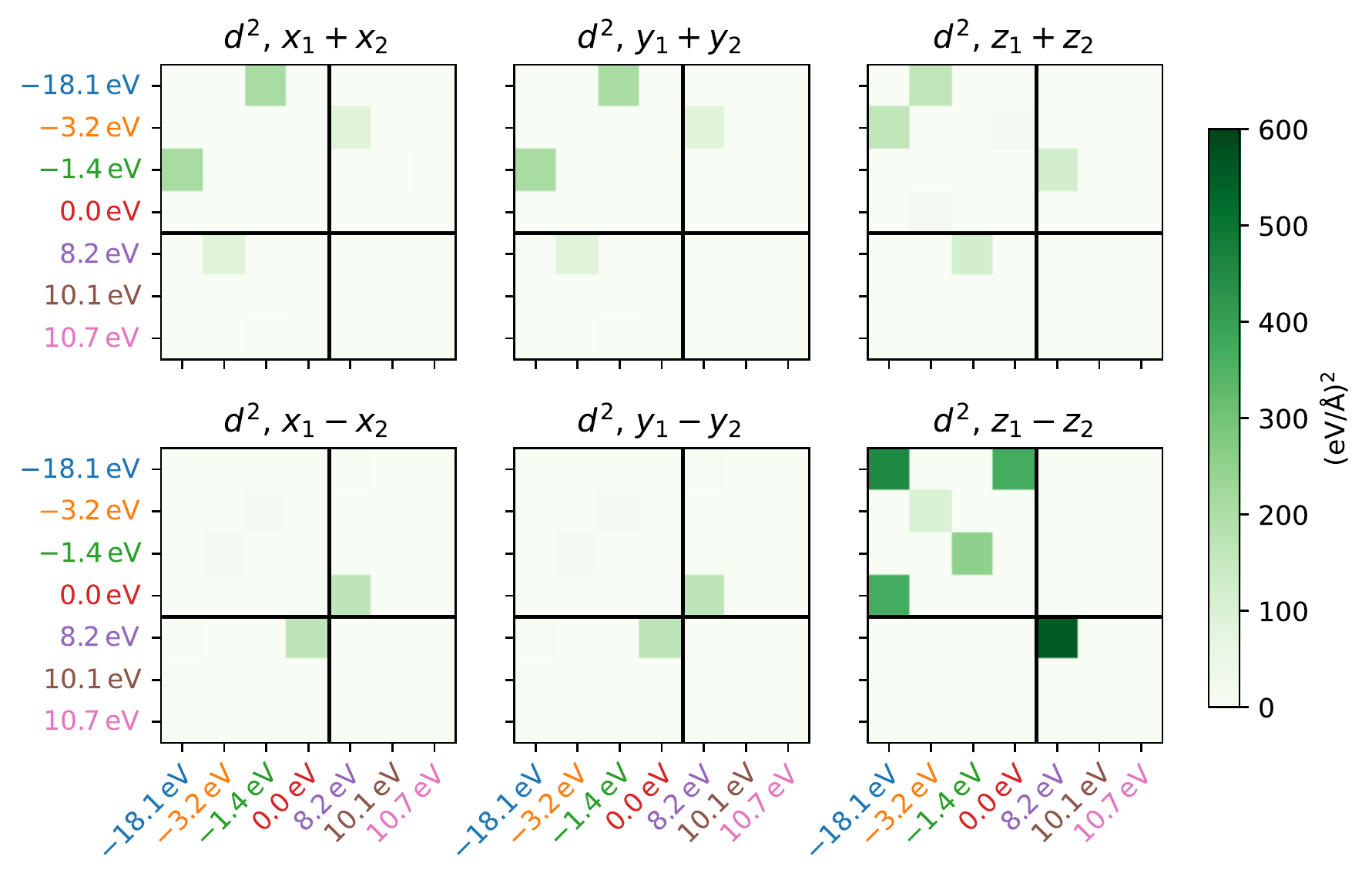}
\caption{Displacement potential $d^2$ in \Ntwo. Every panel stands for the coupling to a single phonon mode; there is no mode-mode coupling in \Ntwo. The colored matrix shows which electronic states couple to the phonon. Degenerate electronic states have been summed over.}
\label{fig:N2:g2}
\end{figure*}

In fluctuation diagnostics, one looks at the contribution of individual electronic states to total change in phonon energy, thereby extracting the relevant screening processes.
Since there is no phonon-mode mixing in \Ntwo\ by symmetry, we can analyze the six phonon-diagonal elements $\Delta \hat{\omega}^2_{\mu\mu}$ one by one.

The first ingredient is the bare susceptibility $\chi^0$, shown in Fig.~\ref{fig:N2:chi0}, quantifying the electronic transitions possible in the system. Due to the Pauli principle, only transitions across the Fermi level are allowed and the susceptibility is largest for pairs of states close to the Fermi level.

In addition to the purely electronic $\chi^0$, there is also the deformation potential.
Figure~\ref{fig:N2:g2} shows $d^2$ (note that this is shorthand for the product of a bare and dressed deformation potential).
We observe that $d^2$ is real, positive, and a symmetric matrix in electronic space. Because of the rotation symmetry, the deformation potential is identical for the $x$ and $y$ phonon modes (note that we have summed over degenerate electronic states).
The magnitude of $d^2$ for the bond-stretching mode is approximately one order of magnitude larger than for the other modes. The deformation potential for this mode is qualitatively different, since it is largely diagonal, whereas the other modes only have off-diagonal elements.
Physically, a diagonal deformation potential means that ionic displacements change the energy of the Kohn-Sham orbitals, whereas off-diagonal elements describe displacement-induced hybridization between orbitals.
Most matrix elements of the electron-phonon coupling are zero, reflecting selection rules.

The total phonon renormalization $m_\mu \Delta \hat{\omega}^2_{\mu\mu}$ is obtained by multiplying $\chi^0$ and $d^2$ and summing over the intermediate electronic states. Fluctuation diagnostics considers these summands one by one.
The results so far show that $\chi^0$ is restricted to transitions across the Fermi level.
The bond-stretching mode has no corresponding electron-phonon coupling, so its energy is barely renormalized.
On the other hand, the five other modes are substantially renormalized. In fact, this renormalization is responsible for breaking Goldstone's theorem.

From Fig.~\ref{fig:N2:g2}, it is clear that only very few combinations of electronic states contribute to the phonon renormalization.
For the three uniform translations, it is worthwhile to analyze in more detail which electronic transitions are responsible for the phonon renormalization and the absence of Goldstone's theorem in the bare phonons.
In all cases, it is the electronic state at 8.2\,eV (LUMO) that is responsible for the screening, combined with either the $-3.2$\,eV or $-1.4$\,eV states, for the in-plane and out-of-plane displacements, respectively. The displacement $x_1+x_2$ is odd under the reflection $x\mapsto -x$, the twofold degenerate 8.2\,eV states are a combination of p$_x$ and p$_y$ orbitals and thus have a p$_x$ component that is odd under this reflection, whereas $-3.2$\,eV is a combination of s and p$_z$ orbitals and is even under the reflection. This makes $\bra{\mathrm p_x} (x_1+x_2)\ket{\mathrm s\mathrm p_z}$ even and the transition is allowed.
Note that $\bra{\mathrm p_x} (x_1+x_2)\ket{\mathrm s\mathrm p_z}$ is also even under the reflections $y\mapsto -y$ and $z\mapsto -z$.
The same argument holds for $y_1+y_2$ and p$_y$.  Finally, for $z_1+z_2$, the state at $-1.4$\,eV is another twofold degenerate combination of p$_x$ and p$_y$ orbitals and the combination $\bra{\mathrm p_x}(z_1+z_2)\ket{\mathrm p_x}$ is even under all three inversions and thus allowed. The structure of the upper panels in Fig.~\ref{fig:N2:g2} reflects the dipole selection rule for uniform translations: Most matrix elements are zero by symmetry.

This also provides a recipe for choosing the electronic target space in such a way that specific partially screened uniform translation modes indeed have zero energy. If the state at $-3.2$\,eV is integrated out and excluded from the constrained space $\Omega'$, then the partially screened phonon modes would already include this coupling. No further coupling to $x_1+x_2$ is dipole allowed, so the partially screened $x_1+x_2$ phonon needs to have zero energy. Similarly, integrating out the $-1.4$\,eV mode would guarantee a zero-energy partially screened $z_1+z_2$ phonon. This construction of Goldstone-preserving target spaces based on dipole selection rules can also be applied to crystalline materials.

Finally, we observe that both $-\chi^0$ and $d^2$ are always positive, so that every summand in the fluctuation diagnostics gives a positive contribution. This is a numerical confirmation of the earlier proof of positivity.

\section{Benzene}
\label{sec:benzene}

\begin{figure}
\includegraphics{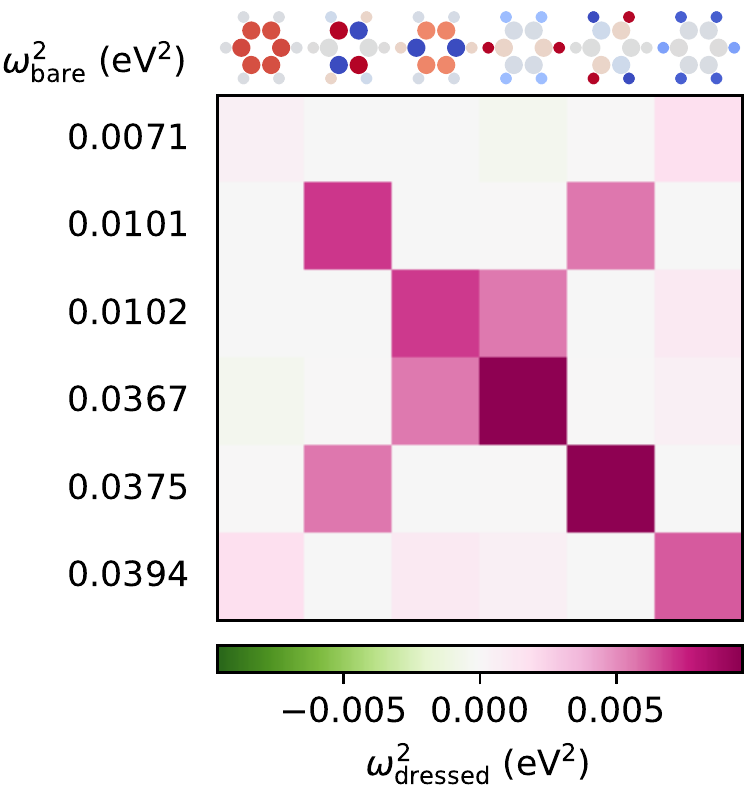}
\caption{Selected phonons in benzene. The dressed (DFPT) dynamical matrix in the basis of bare (cDFPT) eigenmodes. The target subspace spans all bands up to $\pi_6^*$. The dynamical matrix is block diagonal and only a $6 \times 6$ block is shown here (out-of-plane displacements, even under 180-degree rotation symmetry). The bare eigenmodes are shown at the top, with their corresponding energies at the left side of the matrix.}
 \label{fig:C6H6:selectedphonons}
\end{figure}

We now move on to a more complicated molecule, benzene (C$_6$H$_6$). With 12 atoms, this molecule has 36 phonon modes in total.
All atoms lie in a single plane ($z=0$).
By symmetry, in-plane and out-of-plane displacements decouple. We restrict our attention to out-of-plane displacements that are furthermore symmetric under 180-degree rotation around the $z$~axis.
This leaves a subspace consisting of six phonon modes. Figure~\ref{fig:C6H6:selectedphonons} shows the DFPT dynamical matrix $\hat{\omega}^2$ in the basis of cDFPT eigenmodes, where the cDFPT target subspace consists of all electronic eigenstates up to the topmost p$_z$ level ($\pi_6^*$). The presence of off-diagonal elements shows that, unlike in \Ntwo, there is mode-mode coupling. In particular, we find coupling between modes that differ in the relative direction of the C and H atoms~\footnote{In total, there are 12 atoms and thus also 12 out-of-plane modes.
The C$_6$ rotational symmetry divides these up into six pairs of modes, with mode-mode coupling allowed only within these pairs. Three of the six pairs are shown in Fig.~\ref{fig:C6H6:selectedphonons}.}. In the cDFPT eigenbasis, the motion of the C and H ions is almost completely decoupled. It is the electronic chemical bonding that is responsible for the coherent motion of both types of atoms, and this bonding is frozen out in cDFPT\@.
An example of this is the $z$-translation DFPT Goldstone mode. This mode is a linear combination of the cDFPT modes at 0.0071\,eV$^2$ (translation of C atoms) and 0.0394\,eV$^2$ (translation of H atoms).

Comparing the DFPT energies inside the matrix with the cDFPT eigenenergies (to the left) shows that the magnitude of the screening is substantial.
Note that negative off-diagonal elements do appear in the matrix; positivity is only guaranteed for the eigenvalues of $\hat{\omega}^2_\text{cDFPT}$, $\hat{\omega}^2_\text{DFPT}$, and $\hat{\omega}^2_\text{cDFPT}-\hat{\omega}^2_\text{DFPT}$.

\subsection{Partially dressed}
\label{sec:partially_dressed}

\begin{figure}
\includegraphics[width=\linewidth]{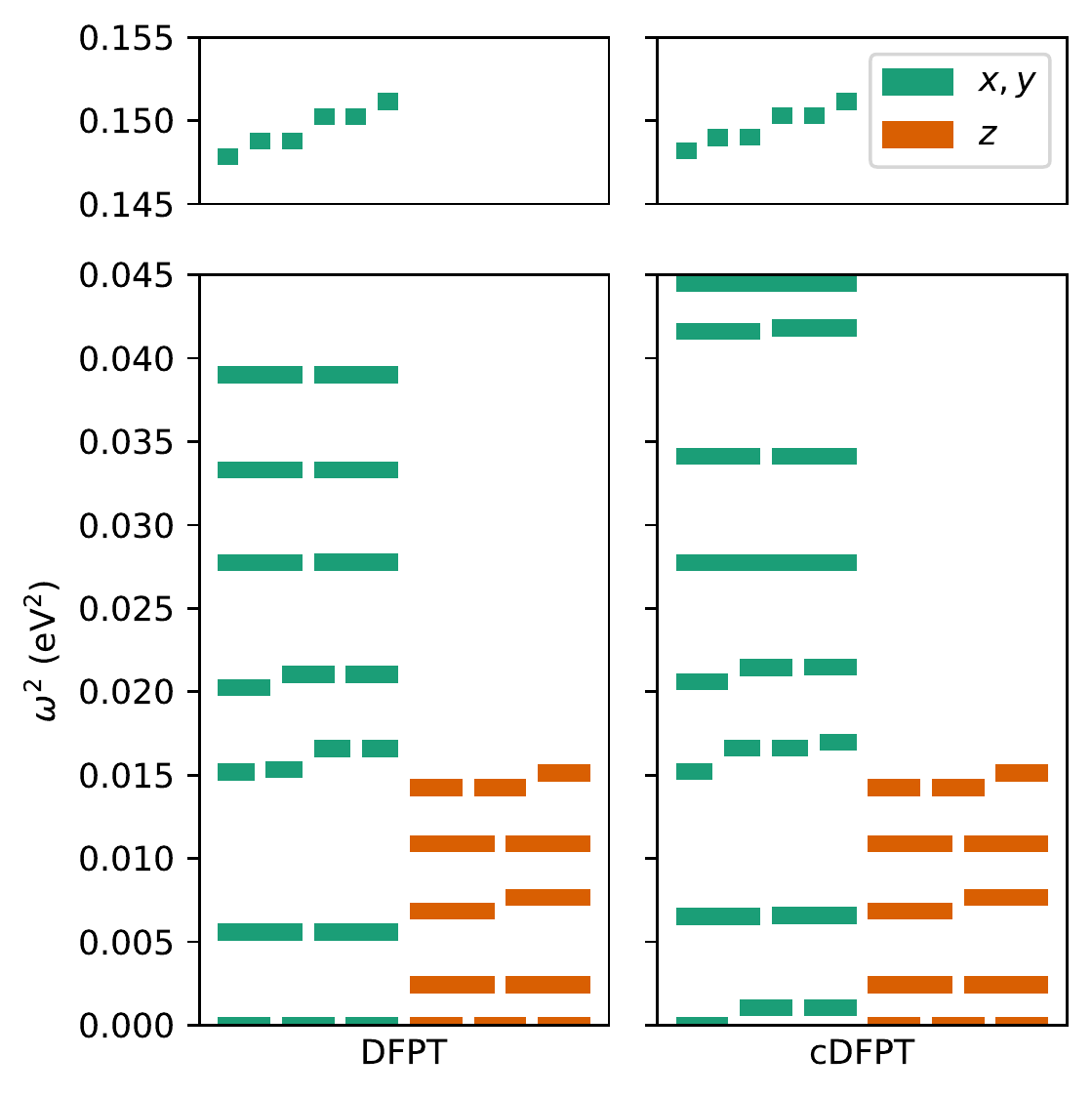}
\caption{DFPT and cDFPT phonons in benzene. $x, y$ ($z$) labels in-plane (out-of-plane) modes. The cDFPT phonons exclude screening from electronic transitions between the six p$_z$~states. The out-of-plane modes are not affected by this constraint due to the mirror-symmetry selection rule.}
 \label{fig:C6H6:phonons}
\end{figure}

So far, we have considered the effect of electronic screening on phonons by either allowing or forbidding screening from all electronic states. The resulting modes are called dressed and bare phonons, respectively. For establishing low-energy models, one is frequently interested in an intermediate object, the partially dressed phonons. In that case, screening by most electrons is allowed and only transitions within a small ``target'' subspace close to the Fermi level are excluded.

\begin{figure*}
 \includegraphics[width=\linewidth/3]{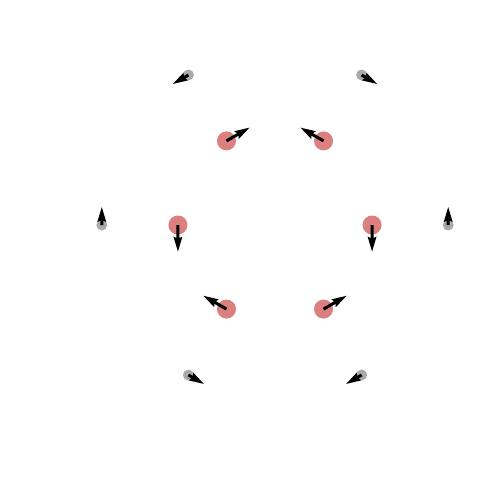}%
 \includegraphics[width=\linewidth/3]{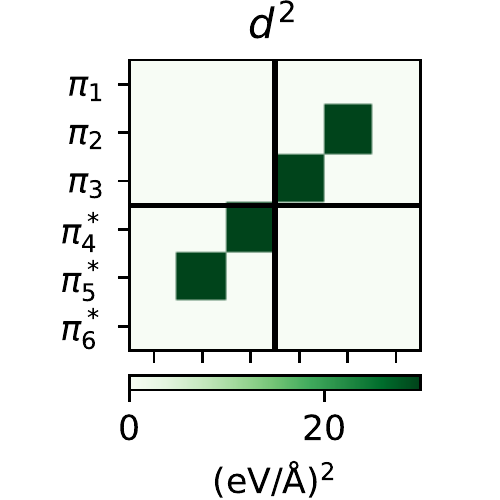}%
 \includegraphics[width=\linewidth/3]{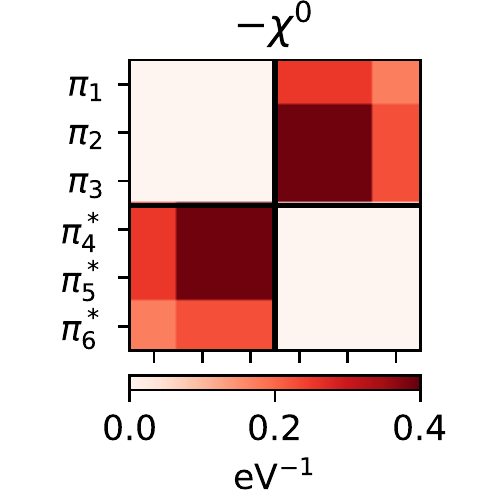}
 \caption{Left: An in-plane phonon (DFPT eigenmode) with substantial coupling to the p$_z$ electrons. Middle: $d^2$ for this mode. Right: Bare electronic susceptibility $\chi^0$. $d^2$ and $\chi^0$ are both visualized in terms of the eigenstates of the electronic p$_z$ space. The contribution to the phonon self-energy is obtained as the \emph{pointwise} product of these two squares.}
 \label{fig:benzene:pzmodes}
\end{figure*}

For benzene, this low-energy subspace is spanned by the six p$_z$ orbitals---analogous to the tight-binding description of graphene---and there is a considerable  Coulomb interaction between these electrons. Minimalist models of benzene consisting of six electronic states therefore regularly feature as a testbed for investigating electron-electron interactions~\cite{Pariser53,Pople55,Valli10,Schuler13,Changlani15,Pudleiner19,intVeld19}. Here, we instead focus on the electron-ion interactions of this subspace.

Figure~\ref{fig:C6H6:phonons} shows the phonon frequencies of both the fully screened DFPT phonons and the partially screened cDFPT phonons. Their difference shows the effect of screening by the p$_z$ orbitals on the phonons. An important observation is that out-of-plane phonons are not screened by the p$_z$ electrons; their frequencies are identical in DFPT and cDFPT\@. This is caused by a symmetry selection rule; all p$_z$ orbitals are odd under mirror symmetry, and this implies $d=0$. An equivalent observation can be made for graphene; see Appendix~\ref{app:graphene}.

By comparing the partially and fully dressed dynamical matrices, we find that the most substantial renormalization occurs for the phonon mode shown in Fig.~\ref{fig:benzene:pzmodes}, an in-plane mode where neighboring C atoms move in opposite directions. The figure also shows the fluctuation diagnostics of this mode. In the middle, the matrix $d^2$ only has contributions coming from excitations between the HOMO and LUMO levels, namely $\pi_2 \leftrightarrow \pi_5^*$ and $\pi_3 \leftrightarrow \pi_4^*$. These are the only combinations of molecular orbitals that are allowed to be coupled to this mode by mirror symmetry. Since these excitations are also energetically favorable for the bare susceptibility $\chi^{0}$ (right), they lead to a large overall renormalization of this phonon mode.

\section{Conclusion}

We have constructed a general framework for \emph{ab initio} downfolding electron-ion problems within the Born-Oppenheimer approximation, and we have shown how the cDFPT fits into this framework.
We have shown analytically that electronic screening lowers the phonon energies. Even the fluctuation diagnostics---the contributions from specific electronic states---are sign definite. The constrained theory can explicitly break symmetries, thereby reducing the number of Goldstone modes. Dipole selection rules can be used to construct a low-energy electronic subspace where even the partially dressed Goldstone modes are guaranteed to have zero energy. We note that the cDFPT phonon dispersion in Fig.~2(a) of Ref.~\citenum{Berges20} indeed satisfies Goldstone's theorem.

We have illustrated the theorem with cDFPT for molecules (nitrogen and benzene), since these are among the simplest and clearest examples of electron-ion coupling. In particular, they provide a clear view on the orbital structure of the theory.

\acknowledgments

The authors would like to thank J\"org Kr\"oger, Susan K\"oppen, Filippo Balzaretti, and Samuel Ponc\'e for useful discussions.
Financial support by the Deutsche Forschungsgemeinschaft (DFG) through GRK 2247 and EXC 2077 and computational resources of the North-German Supercomputing Alliance (HLRN) are gratefully acknowledged.
EvL is supported by the Central Research Development Fund of the University of Bremen.

\appendix

\section{Manifolds and derivatives}
\label{app:kohnsham}

In DFPT, the electronic configuration is described by the density operator $\hat{\rho}$. The Kohn-Sham electronic energy functional is $\mathcal{E}(\hat{\rho}) = \Tr \hat{\rho} H$, where $H$ is the Kohn-Sham Hamiltonian. Let us initially assume that $H$ is independent of $\hat{\rho}$. One could anticipate that the Hessian matrix, the second derivative of $\mathcal{E}$ with respect to $\hat{\rho}$, would be zero. However, at this point it is important to realize that the derivatives are taken on the manifold of density matrices and the Hessian on a curved manifold contains additional terms.

To see how this works, we consider a minimal system of two electronic levels with single-particle energies $\epsilon_a < \epsilon_b$ and corresponding states $\ket{a}$, $\ket{b}$. The Hamiltonian is
$
 H = \epsilon_a \ketbra{a}{a} +\epsilon_b \ketbra{b}{b}.
$
The ground state is given by the ground-state density operator $\hat{\rho}_0$. If $\epsilon_a$ and $\epsilon_b$ are on the same side of the Fermi level, the system is completely filled or empty. The most interesting situation occurs when $\epsilon_a<E_f<\epsilon_b$. In that case, the ground-state $\hat{\rho}_0 = \ketbra{a}{a}$ is the projection operator onto $a$ and the total number of electrons is 1.
The manifold of allowed density matrices is given by all density matrices with total density 1. It is easy to see that $\hat{\rho}_\theta = (\cos \theta\ket{ a}+\sin \theta\ket{ b})(\cos \theta\bra{ a}+\sin \theta\bra{ b})$ lies in this manifold for any $\theta$. The corresponding energy is
\begin{align}
 \mathcal{E}(\hat{\rho}_\theta) &= \Tr \hat{\rho}_\theta H \\
 &= \epsilon_a \cos^2 \theta + \epsilon_b \sin^2 \theta,
\end{align}
which has first derivative zero at $\theta=0$, and the second derivative at $\theta=0$ is
\begin{align}
 \frac{d^2\mathcal{E}}{d\theta}\Big|_{\theta=0}
 &= 2 (\epsilon_b-\epsilon_a)=2 (\hat{\chi}^0_{ab,ab})^{-1}  \label{eq:lindhard}\\
 &=  \frac{d^2\mathcal{E}}{d\hat{\rho}_{ab}^2}+\frac{d^2\mathcal{E}}{d\hat{\rho}_{ba}^2}.
\end{align}
The inverse bare susceptibility at $T=0$ in Eq.~\eqref{eq:lindhard} enters the phonon self-energy.
Generalizing the result to higher-dimensional spaces shows that off-diagonal terms $\partial_{ab}\partial_{cd}$ vanish, so $\hat{\chi}^0_{ab,cd}$ is a diagonal matrix, as stated in the main text.

So far, we have assumed that the Kohn-Sham Hamiltonian $H$ is independent of $\hat{\rho}$. In reality, DFT is a self-consistent theory and the Kohn-Sham Hamiltonian contains the Hartree and exchange-correlation potentials. It is generally possible to write
\begin{align}
  \frac{d^2\mathcal{E}}{d\hat{\rho}_{\alpha\beta}d\hat{\rho}_{\gamma\delta}} \Big|_{\hat{\rho}_0}
  &= (\hat{\chi}^0)^{-1}_{\alpha\beta,\gamma\delta} - \hat{U}_{\alpha\beta,\gamma\delta},
\end{align}
which here merely acts as the definition of $\hat{U}$. $\hat{U}$ accounts for the electron-electron interactions that are not present in the auxiliary Kohn-Sham system. In cDFPT calculations, $\hat{U}$ is formally incorporated into $\tilde{d}$ (see Fig.~\ref{fig:vertex}) and never calculated explicitly. Still, this formal relation is necessary for some of the proofs in the main text.

\begin{figure}
\includegraphics[width=\linewidth]{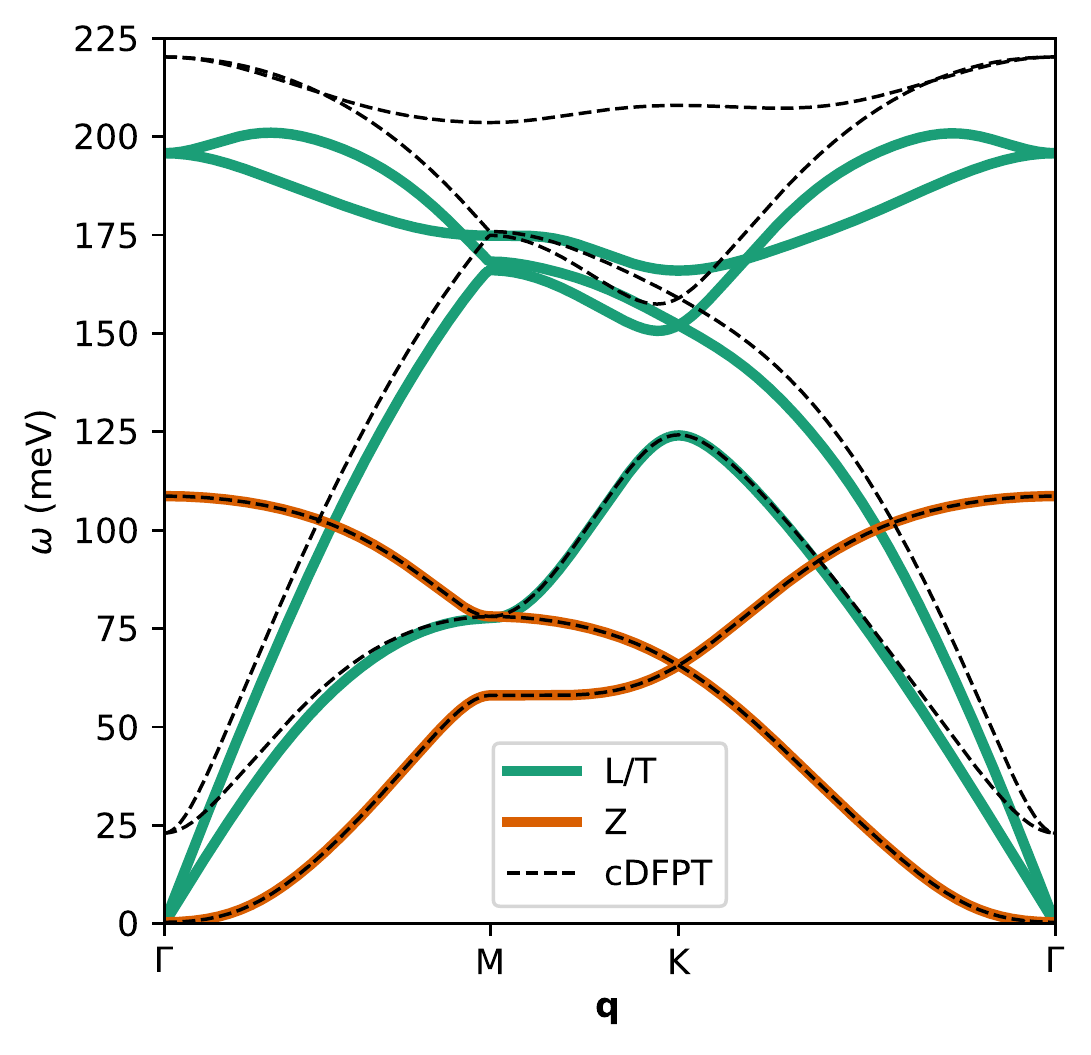}
\caption{DFPT and cDFPT phonons of graphene. L/T (Z) labels longitudinal/transverse in-plane (out-of-plane) modes from DFPT\@. The cDFPT phonons exclude electronic screening from within the two p$_z$~bands. Please note that for the out-of-plane modes, we Fourier interpolated $\hat{\omega}_{\mathbf{q}}$ instead of $\hat{\omega}^2_{\mathbf{q}}$.}
 \label{fig:graphene}
\end{figure}

\section{Sign of phonon self-energy}
\label{app:lemma}

From the definition, $\hat{\chi}^0$ is diagonal (as a $N_\text{el}^2 \times N_\text{el}^2$ matrix) and thus also symmetric. The diagonal elements are given by the Lindhard expression and are negative or zero, since $f$ is a decreasing function. Thus, $-\hat{\chi}^0$ is positive semidefinite. The matrix $\hat{U}$ is real and symmetric. From this, it follows that $\hat{\chi}=-\hat{\chi}^0/(\mathbb{I}-\hat{U}\hat{\chi}^0)$ is also real and symmetric, which can be checked order by order as $(\hat{\chi}^0 \hat{U} \hat{\chi}^0 \ldots \hat{\chi}^0 \hat{U} \hat{\chi}^0)^T=(\hat{\chi}^0)^T \hat{U}^T (\hat{\chi}^0)^T \ldots (\hat{\chi}^0)^T \hat{U}^T (\hat{\chi}^0)^T = \hat{\chi}^0 \hat{U} \hat{\chi}^0 \ldots \hat{\chi}^0 \hat{U} \hat{\chi}^0$. As long as the geometric series is convergent, which corresponds to thermodynamic stability, it does not change the sign of eigenvalues and $\hat{\chi}$ will be positive semidefinite and symmetric.

A further detail that is necessary for our proof: $P$ is an orthogonal projection with respect to the original inner product but not with respect to the inner product defined by $\hat{\chi}$, so it is not immediately trivial that $\langle d,Pd\rangle^\chi \geq 0$. Below, $\hat{\chi}$ is written without hat for convenience. Since $P$ is a projection operator, we can write $d=\alpha+\beta$ with $\alpha \equiv Pd$, $P\alpha=P^2 d = Pd=\alpha$ and $\beta \equiv d-\alpha$, $P\beta=Pd-P\alpha = \alpha-\alpha=0$.

 We wish to show that
 \begin{align}
  d^T \chi P d = \alpha^T \chi \alpha + \beta^T \chi \alpha \overset{?}{\geq} 0.
 \end{align}
If $\beta^T \chi \alpha>0$, this is proven immediately, since $\alpha^T \chi \alpha >0$.
Otherwise, the trick is to take $x = \alpha + \lambda \beta $ with a real number $\lambda$ so that
\begin{multline}
 0\leq x^T \chi x = \alpha^T \chi \alpha + 2\lambda \beta^T \chi \alpha + \lambda^2 \beta^T \chi \beta \\
 = \alpha^T \chi \alpha + \beta^T \chi \alpha.
\end{multline}
The last equality is used to solve for $\lambda$ and a real solution $\lambda$ can be found as long as the discriminant $D/4 = (\beta^T \chi \alpha)^2 - (\beta^T \chi \alpha)(\beta^T \chi \beta)$ is positive. But we were studying the case $\beta^T \chi \alpha < 0$, so indeed $D>0$, $\lambda$ can be chosen appropriately, and the proof of the lemma is finished.

\hfill

\section{Graphene}
\label{app:graphene}

To illustrate how the presented results for molecules can be transferred to periodic systems, we consider the example of graphene. Following up on Sec.~\ref{sec:partially_dressed}, Fig.~\ref{fig:graphene} shows the phonon dispersion of graphene from DFPT (colored solid lines) and cDFPT (black dashed lines). We choose as the target subspace the two p$_z$ bands forming the Dirac cones at the K points. As in Fig.~\ref{fig:C6H6:phonons}, excluding the electronic screening from within the p$_z$ manifold does not alter the out-of-plane modes (orange) at all. By contrast, the frequencies of the in-plane modes (green) are increased, even in the case of two acoustic modes at~$\Gamma$. Again, these partially screened phonons do not satisfy Goldstone's theorem.

\bibliography{references}

\end{document}